# In-situ Instrumental Setup for Influence Study of Hard-axis Bias Magnetic Field on MR transfer curves of sing MTJ sensor and MTJs array sensor

L. Li[1], Y. Zhou[2], W.W. Wang[3], P. W. T. Pong[1]

[1]Department of Electrical and Electronic Engineering, The University of Hong Kong,
[2]Department of Physics, The University of Hong Kong, Hong Kong
[3]Department of Physics, Ningbo University, Ningbo 315211, China

**Abstract**

Establishment of home-made measurement setups for the characterization of MR sensor is proposed and described here. The MR loops of MR sensors can be obtained with the instrument using two-point probe measurement and four-point probe measurement. Two pairs of Helmholtz coils can supply a hard-axis magnetic field and a soft-axis magnetic field for the sensor. The single MTJ sensor and MTJs array sensor in Wheatstone bridge were characterized and compared here. The influence of hard-bias magnetic field on MR transfer curves of sing MTJ sensor and MTJs array sensor are investigated. The corresponding optimal hard-axis magnetic fields were obtained through Helmholtz coils to eliminate the hysteresis for linear response of single MTJ sensor and MTJs array sensor.

**Introduction**

The sensing of magnetic field has been important to human life for over hundreds of years since the compass with loadstone was used for navigation. Generally speaking, magnetic field sensor is a kind of device with characteristics changing as a function of an external magnetic field. With the development of highly-sensitive magnetic field sensors based on magnetoresistance (MR) effect in recent years, the MR-based magnetic field sensors have been successfully applied in hard disk drives and magnetic memories, which offer an inspiration for their use in magnetic biodetection. Typical MR sensors include GMR sensors and magnetic tunnel junction (MTJ) sensors. The magnetic biosensor system utilizing GMR sensor called BARC is firstly demonstrated by a group at the Naval Research Laboratory and NVE Corporation [1, 2]. Since then, it has become a much sought-after research topic in recent years, particularly with the advances in nanofabrication techniques. Moreover, the nature of solid-state thin film sensors makes the MR sensors can be integrated into a high-density MR array sensor, which is well suited for multiplex biodetection.[3]

Before their application in our biodetection platform, it is important to carry out the characterizations of the MR sensors. Here, we propose an in-situ instrumental setup for characterizing the MR sensors and investigate the influence of hard-axis bias magnetic field on the MR transfer curves of single MTJ sensor and MTJs array sensor.

**Experimental setup**

**Two-point probe measurement**

The simplest way to characterize the MR sensor is to apply a constant current across the two terminals of the sensor, and measure the corresponding voltage. Alternately, a constant voltage can be supplied and the current is measured to obtain the sensor resistance. The traditional two-point probe is illustrated in Fig. 1. When the MR sensor is supplied with a constant current I by current source meter, the measured resistance $R_M$ through two-point probe sensing, and is given by $R_M = \frac{V_M}{I}$, where $V_M$ is voltage measured experimentally. However, the resistance $R_s$ of MR sensor is given by $R_s = \frac{V_R}{I}$, and $V_R$ is the voltage across MR sensor. Thus, $R_M = R_s + 2R_{LEAD}$, where $R_{LEAD}$ is the lead resistance with contact resistance, that is to say, the measured resistance $R_M$ is a little larger than the actual resistance of MR sensor $R_s$.

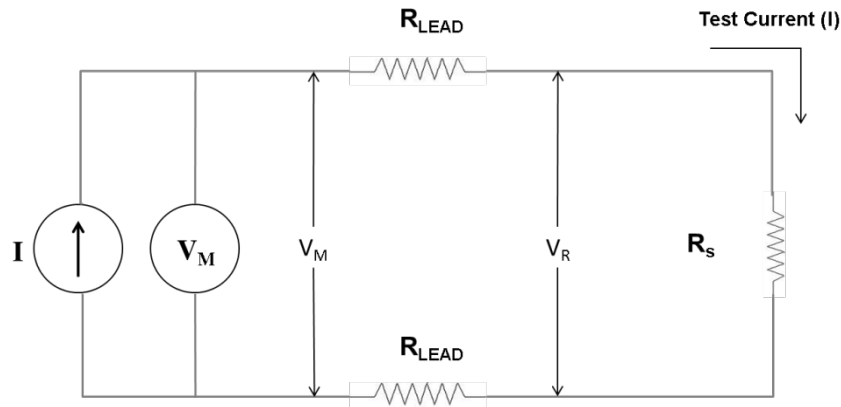

Figure 1 Two-point probe resistance measurement for MR sensor.

As shown in Fig. 2, the two-point probe measuring setup consists of a PC computer with LabVIEW program and PCI-GPIB interface, Keithley 2400 source meter, Keithley 2000 digital multimeter, two sets of Kikusui Bipolar Power Supply PBX 40-5 and two pairs of home-made Helmholtz coils. The Keithley 2400 source meter served to provide biasing current or voltage for MR sensor and the Keithley 2000 digital multimeter is used to measure the output signal. The Helmholtz coils were supplied by the Kikusui Bipolar Power Supply PBX 40-5 to provide magnetic field ranging from -200 Oe to +200 Oe. For the magnetically anisotropic MR sensors, the magnetic field applied in the easy-axis direction of MR sensor provides the scanning fields for MR-loop measurement and the magnetic field applied in hard-axis direction eliminates the hysteresis to increase the linear response of MR sensor [4]. All the instruments were connected to the PC computer via GPIB cables, and can be controlled through LabVIEW program.

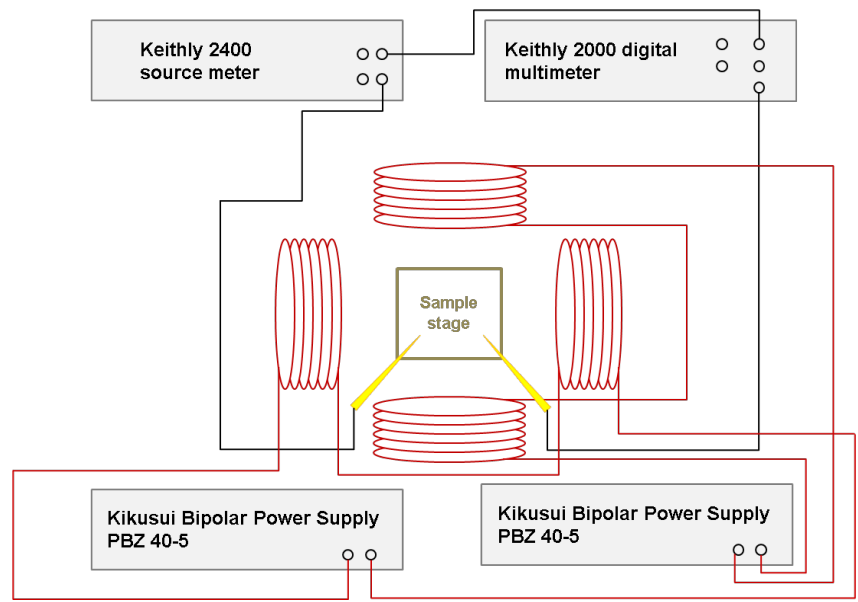

Figure 2 Schematic diagram of the instrumental setup for characterization of MR array sensor. The MR array sensor is mounted on the sample stage and measured by two-point probe electrical method.

**Four-point probe measurement**

For the MR sensor with four contact pad, the characterization can be performed through a four-point probe electrical method. Four-point probe electrical method that uses separate pairs of current-carrying and voltage-sensing electrode can make more accurate resistance measurement than traditional two-probe electrical method. As illustrated in Fig. 3, the resistance $R_s$ of MR sensor can be measured through four-point probe sensing, and is given by $R_s = \frac{V_M}{I} = \frac{V_R}{I}$, where I is test current supplied to MR sensor by current source meter, $V_M$ is voltage measured experimentally, and $V_R$ is the voltage across MR sensor. $V_M = V_R$ because sense current is negligible.

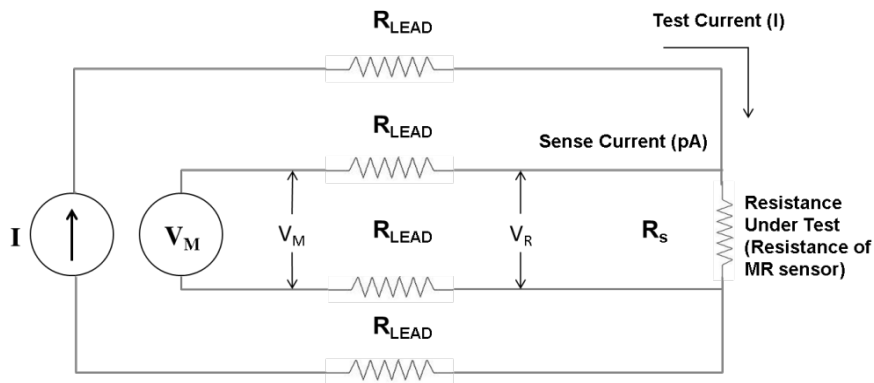

Figure 3 Four-point probe resistance measurement for MR sensor.

As shown in Fig. 4, the four-point probe electrical measuring setup consists of a PC computer with LabVIEW program and PCI-GPIB interface, Keithley 2400 source meter, Keithley 2000 digital multimeter, two sets of Kikusui Bipolar Power Supply PBX 40-5 and two pairs of home-made Helmholtz coils. The Keithley 2400 source meter served to provide biasing current or voltage for MR sensor and the Keithley 2000 digital multimeter is used to measure the voltage. The Helmholtz coils were supplied by the Kikusui Bipolar Power Supply PBX 40-5 and they can provide magnetic field ranging from -200 Oe to +200 Oe. The magnetic field applied in the easy-axis direction of MR sensor provides the scanning fields for MR-loop measurement and the magnetic field applied in hard-axis direction eliminates the hysteresis to increase the linear response of MR sensor [4]. All the instruments were connected to the PC computer via GPIB cables, and can be controlled through LabVIEW program.

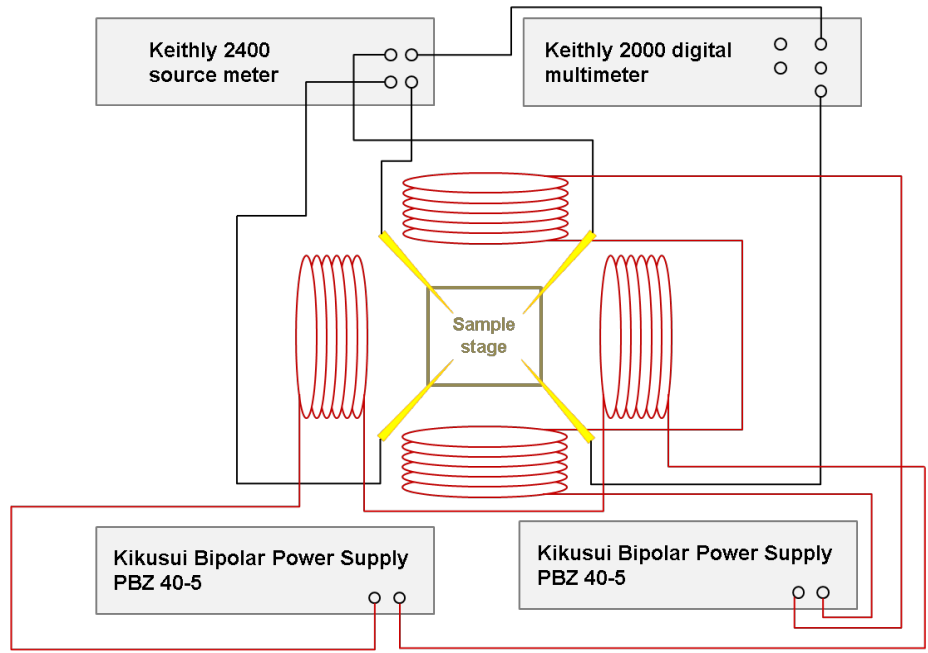

Figure 4 Schematic diagram of the instrumental setup for characterization of single MR sensor. The MR sensor is mounted on the sample stage and measured by four-point probe electrical method.

**Development of LabVIEW Software**

The whole system and data acquisition process are controlled by a program using LabVIEW software. The interface of the LabVIEW control program is illustrated in Fig. 5. The LabVIEW program control the MR sensor characterization system, acquires the data, processes the data, and displays them in real-time charts on PC computer monitor. It also stores the operation parameters and output signals as data files for further processing by using software such as Matlab, Excel, or Origin. The LabVIEW software programmed here was based on the work of Raymond's version.

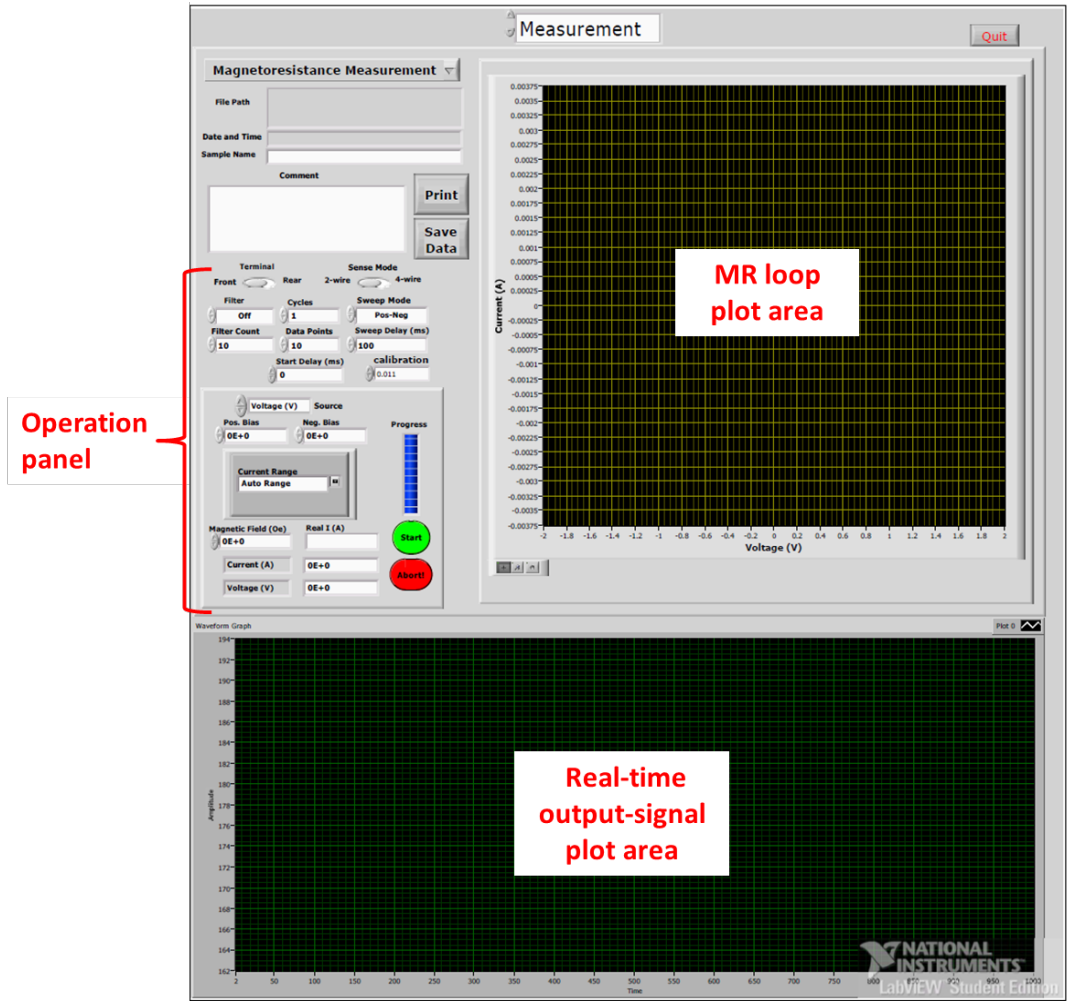

Figure 5 The interface of LabVIEW program for MR measurement. On the upper left is the operation panel, on the upper right is the MR loop plot area, and on the lower part is the real-time output-signal plot area.

## Results and discussion

### Single MTJ sensor

The single MR sensor characterized here is a MTJ sample purchased from Micro Magnetics. It was prepared by magnetron sputtering on thermally oxidized silicon wafers with a base pressure of $2\times10^{-10}$ Torr. The thin-film stack structure was (units in nanometer): substrate/Ta (5)/Ru (30)/Ta (5)/CoFe (2)/IrMn (15)/CoFe (2)/Ru (0.8)/CoFeB (3)/MgO (1.9)/CoFeB (3)/Ta (5)/Ru (10). The working (sensing) area was $4\times2$ $\mu m^2$ and patterned by optical lithography and ion milling. A thickness of 200-nm gold layer was deposited on the top of MTJ sensing area. The MTJ sensor was observed under SEM imaging as shown in Fig. 6. There are four gold pads (as indicated by yellow words) which can be utilized as wire-bond pads to carry out the standard four-point probe measurement for the MR transfer curve of sensor. The MTJ sensor placed on sample stage were shielded in a mu-metal shielding box, avoiding the disturbances of ambient magnetic field. Two pairs of Helmholtz coils were utilized to provide the magnetic field along the easy-axis and hard-axis of the MTJ sensor, respectively. All the measurements were carried out at room temperature.

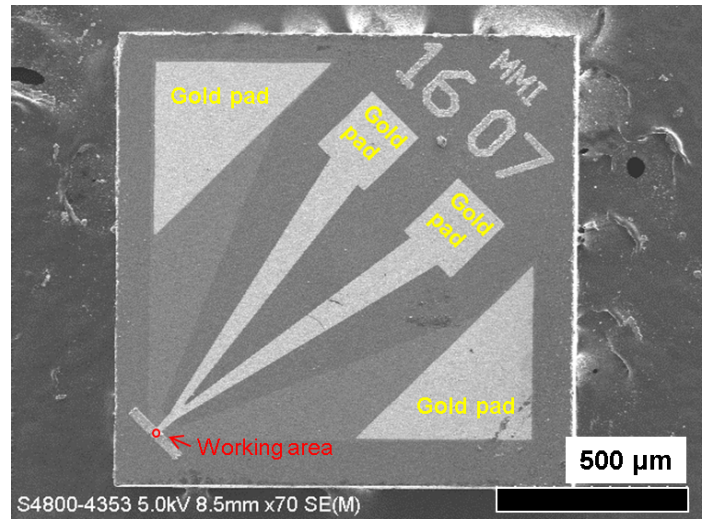

Figure 6 SEM image of single MTJ sensor. Four gold pads (wire-bond pads) are labeled by yellow words. Working area of the sensor is indicated by red circle.

The MR loop measurement result with an external magnetic field applied along the easy-axis in absence of any hard-axis bias field is shown in Fig. 7 The TMR ratio is 99.1 % and the easy-axis coercivity is 1.5 Oe. The hysteresis was removed by applying a hard axis bias magnetic field, subsequently.

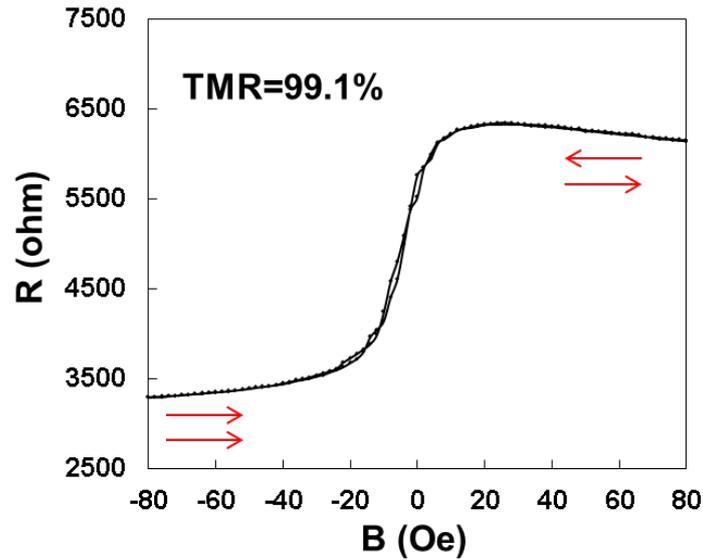

Figure 7 MR loop of MTJ sensor. The MTJ exhibited TMR of 99.1 % in absence of hard-axis bias field.

In Fig. 8, it shows the influence of hard-axis bias magnetic field on the MR transfer curves of MTJ sensor. As the hard-axis bias magnetic field increased from 10 Oe to 30 Oe, the hysteresis reduced. The sensor hysteresis was eliminated at a hard-axis bias field of 60 Oe, meanwhile the sensor sensitivity was found to be 0.82%/Oe from the slope of the transfer curve at low field. The elimination of hysteresis can be interpreted by Stoner-Wohlfarth (SW) model [5]. As the hard-axis bias field further increased from 60 Oe to 80 Oe, the slope of the transfer curve became less steep and thus the sensor sensitivity decreased. Thus, a hard-axis bias magnetic field of 60 Oe is suitable to be used to eliminate the hysteresis of the MR curve while keeping its relatively high sensing sensitivity.

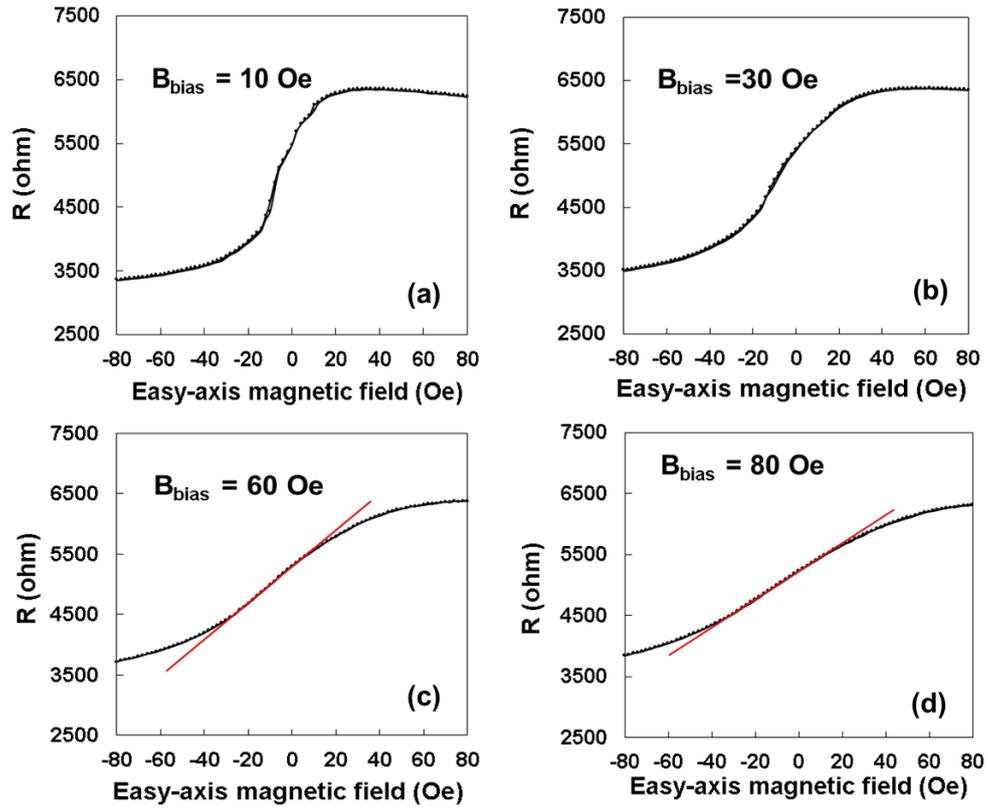

Figure 8 MR loops of MTJ sensor with different biasing fields along the hard-axis. From (a), and (b), we can see the hysteresis reduced with the hard-axis bias field. (c) The hysteresis is removed at 60 Oe of hard-axis bias field and the sensitivity of sensor is 0.82%/Oe. (d) Further increase of hard-axis bias field to 80 Oe reduces TMR ratio and the slope of the linear region of sensor.

**MTJs array sensor**

The MR array sensor characterized here is MTJs array sample purchased from Micro Magnetics. It consists of 76 junctions, and every junction stack was prepared through similar procedure as the single MTJ stack described previously. Micron-size (20 ×120) μm$^2$ elliptical junctions were patterned using standard optical lithography and ion milling. The working (sensing) area of whole sensor was 890 ×890 μm$^2$, which is almost equal to the whole surface area of sensor. The MTJs array sensor was observed under SEM imaging as shown in Fig. 9. There are two gold pads (as indicated by red words) which can be utilized as wire-bond pads to carry out the traditional two-point probe measurement for the MR transfer curve of sensor. The MTJs array sensor was placed on sample stage, shielded in a mu-metal shielding box to avoid the disturbances of ambient magnetic field. Two pairs of Helmholtz coils were utilized to provide the magnetic field along the easy-axis and hard-axis of the MTJs array sensor, respectively. All the measurements were carried out at room temperature.

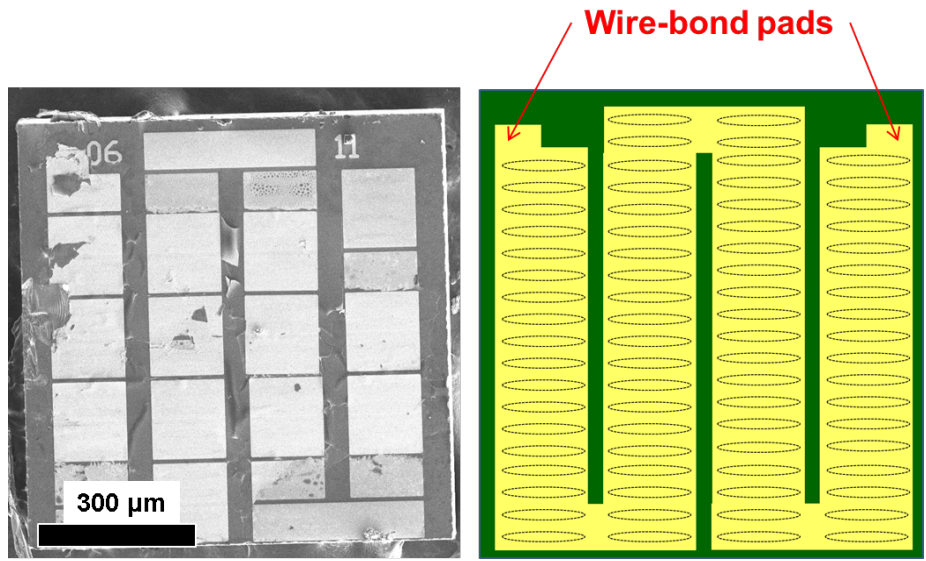

Figure 9 (a) SEM image and (b) schematic drawing of MTJs array sensor. Two wire-bond pads are labeled by red words.

In Fig. 10, it shows the influence of hard-axis bias magnetic field on the MR transfer curves of MTJs array sensor. The MR loop measurement result with an external magnetic field applied along the easy-axis in absence of any hard-axis bias field is shown in Fig. 10a. The TMR ratio is 106 % and the easy-axis coercivity is 1 Oe. The hysteresis was removed by applying a hard axis bias magnetic field of 10 Oe. With a 10-Oe hard-axis bias magnetic field, the sensor sensitivity was found to be 1.39%/Oe from the slope of the transfer curve at low field. As the hard-axis bias field further increased from 30 Oe to 60 Oe, the slope of the transfer curve became less steep and thus the sensor sensitivity decreased. Thus, a hard-axis bias magnetic field of 10 Oe is suitable to be used to eliminate the hysteresis of the MR curve while keeping the relatively high sensing sensitivity of MTJs array sensor.

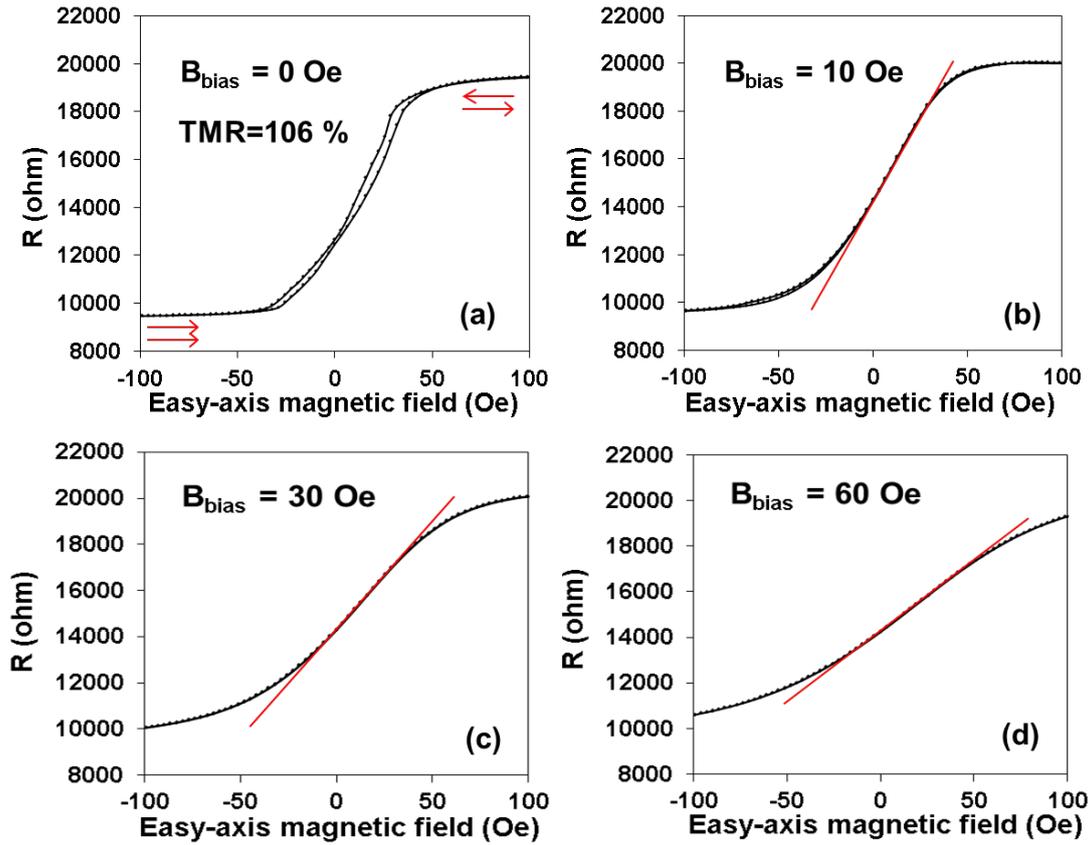

Figure 10 MR loops of MTJs array sensor with different biasing fields along the hard-axis. From (a), and (b), we can see the hysteresis reduced with the hard-axis bias field. The hysteresis is removed at 10 Oe of hard-axis bias field and the sensitivity of sensor is 1.39%/Oe. (d) and (e) show the further increase of hard-axis bias field from 30 Oe to 60 Oe reduces TMR ratio and the slope of the linear region of sensor.

**Conclusion**

We established the setups for the characterization of MR sensor. The setup and data acquisition is controlled by a computer through a LabVIEW program. The MR loops of MR sensors can be obtained with the instrument through two-point probe measurement, four-point probe measurement, or Wheatstone bridge measurement. A hard-axis magnetic field can be applied through Helmholtz coils to eliminate the hysteresis for linear response of the sensor. The measurement results show that out measurement setup is reliable and effective for characterizing the single MTJ sensor, and MTJs array sensor.